\documentclass[a4paper,aps,twocolumn,nofootinbib]{revtex4}
\RequirePackage[colorlinks,hyperindex]{hyperref}
\RequirePackage[english]{babel}
\RequirePackage[latin1]{inputenc}
\RequirePackage[T1]{fontenc}
\RequirePackage{mathrsfs}
\RequirePackage{amsmath}
\RequirePackage{amssymb}
\RequirePackage{amsbsy}
\RequirePackage{color}
\RequirePackage{bm}
\hypersetup{colorlinks=true,breaklinks=true,urlcolor=blue,linkcolor=red}
\begin{document}
\title{\bf{Necessity of Tensorial Connections for Spinorial Systems}}
\author{Luca Fabbri\footnote{fabbri@dime.unige.it}}
\affiliation{DIME, Sez. Metodi e Modelli Matematici,\\
Universit\`{a} di Genova,\\
Via all'Opera Pia 15,\\
16145 Genova, ITALY}
\date{\today}
\begin{abstract}
We consider spinorial fields in polar form to deduce their respective tensorial connection in various physical situations: we show that in some cases the tensorial connection is a useful tool, instead in other cases it arises as a necessary object. The comparative analysis of the different cases possessing a tensorial connection is done, investigating the analogies between space-time structures. Eventual comments on quantum field theory and specific spinors are given.
\end{abstract}
\maketitle
\section{Introduction}
In the study of spinor fields \cite{L, Cavalcanti:2014wia, HoffdaSilva:2017waf, daSilva:2012wp, Ablamowicz:2014rpa, Rodrigues:2005yz, Vignolo:2011qt, daRocha:2013qhu, Ahluwalia:2004ab, Ahluwalia:2004sz, Ahluwalia:2016jwz, Ahluwalia:2016rwl, daRocha:2007pz, HoffdaSilva:2009is, daRocha:2008we, Fabbri:2010ws}, one important way to examine their character is through the so-called \emph{polar form}, the form where each spinor component is written as a product between a module and a phase while respecting manifest covariance \cite{Fabbri:2016msm, Fabbri:2018crr, Fabbri:2016laz, Fabbri:2019kfr, Fabbri:2020ypd, Fabbri:2021mfc, Fabbri:2020elt, Fabbri:2017xyk, Fabbri:2019vut}. There are many motivations to write spinor fields in polar form: a first is that in polar form the spinor field theory can be re-formulated in a way that is independent on the specific representation of the gamma matrices and in terms of only real tensors; then, a second advantage is that in polar variables, it becomes easy to identify, within the spinor field, its $2$ true degrees of freedom and have them isolated from its $6$ Goldstone states \cite{Fabbri:2021mfc}. These Goldstone states are the components that can be transferred into the frame, where they combine with the underlying connection to form objects that, while containing the same information of the connection itself, are nevertheless real tensors, called \emph{tensorial connection} \cite{Fabbri:2018crr,Fabbri:2019kfr}. Writing spinor fields in polar form, and hence being able to compute their tensorial connection, is the third advantage of this formulation, because the tensorial connection turns out to be important, and in some cases even necessary, for spinor fields and their dynamical character, as we are going to show in the following.

The existence of such tensorial connection is a mathematical generality. However, in trivial space-times, using plane waves results into a zero tensorial connection, with the consequence that nowhere in Quantum Field Theory is the tensorial connection ever employed. Just the same, there are important physical systems, like for example in hydrogen-like atoms, where the tensorial connection can not be assumed to vanish \cite{Fabbri:2018crr}. Both mathematical generality and physical applications suggest that the tensorial connection can not be neglected. In the following we are going to consider an example of field, that is the neutral spinor (or Majorana) field, showing that in this case setting to zero the tensorial connection would simply result into a contradiction. We are also going to discuss what is the implication of this fact for the existence of a recently- introduced type of spinor field known as ELKO.
\section{Generalities on Spinorial Fields}
\label{A}
We begin the presentation by recalling the fundamental definition of spinor fields: as spinor fields are defined in terms of their transformation, it is important to show what the general form of this transformation is. To do so, we start by introducing the basis of the Clifford algebra.

In the most general form complex Lorentz transformations are given in terms of Clifford matrices $\boldsymbol{\gamma}_{a}$ such that 
\begin{eqnarray}
&\{\boldsymbol{\gamma}_{a},\!\boldsymbol{\gamma}_{b}\}\!=\!2\mathbb{I}\eta_{ab}
\end{eqnarray}
where $\eta_{ab}$ is the Minkowskian matrix. Defining
\begin{eqnarray}
&\frac{1}{4}\left[\boldsymbol{\gamma}_{a},\!\boldsymbol{\gamma}_{b}\right]
\!=\!\boldsymbol{\sigma}_{ab}
\end{eqnarray}
it is possible to see that $\boldsymbol{\sigma}_{ab}$ are the complex generators of the Lorentz transformations. Because $\boldsymbol{\sigma}_{ab}$ also verify
\begin{eqnarray}
&2i\boldsymbol{\sigma}_{ab}\!=\!\varepsilon_{abcd}\boldsymbol{\pi}\boldsymbol{\sigma}^{cd}
\end{eqnarray}
implicitly defining the $\boldsymbol{\pi}$ matrix (this matrix is normally denoted as a gamma with an index five, but because in the space-time this index has no meaning we will employ a notation with no index), we can infer that the complex Lorentz transformation is reducible. Additionally
\begin{eqnarray}
&\boldsymbol{\gamma}_{i}\boldsymbol{\gamma}_{j}\boldsymbol{\gamma}_{k}
\!=\!\boldsymbol{\gamma}_{i}\eta_{jk}-\boldsymbol{\gamma}_{j}\eta_{ik}
\!+\!\boldsymbol{\gamma}_{k}\eta_{ij}
\!+\!i\varepsilon_{ijkq}\boldsymbol{\pi}\boldsymbol{\gamma}^{q}
\end{eqnarray}
is valid as a geometric identity.

With the $\boldsymbol{\sigma}_{ab}$ and parameters $\theta_{ij}\!=\!-\theta_{ji}$ we can write 
\begin{eqnarray}
&\boldsymbol{\Lambda}\!=\!e^{\frac{1}{2}\theta_{ab}\boldsymbol{\sigma}^{ab}}
\end{eqnarray}
as the complex Lorentz transformations in the most general case. It is possible to make this explicit by defining
\begin{eqnarray}
a\!=\!-\frac{1}{8}\theta_{ij}\theta^{ij}\\
b\!=\!\frac{1}{16}\theta_{ij}\theta_{ab}\varepsilon^{ijab}
\end{eqnarray}
and then
\begin{eqnarray}
2x^{2}\!=\!a\!+\!\sqrt{a^{2}\!+\!b^{2}}\\
2y^{2}\!=\!-a\!+\!\sqrt{a^{2}\!+\!b^{2}}
\end{eqnarray}
so to introduce
\begin{eqnarray}
\cos{y}\cosh{x}\!=\!X\\
\sin{y}\sinh{x}\!=\!Y\\
\nonumber
\left(\frac{x\sinh{x}\cos{y}
+y\sin{y}\cosh{x}}{x^{2}+y^{2}}\right)\theta^{ab}+\\
+\left(\frac{x\cosh{x}\sin{y}
-y\cos{y}\sinh{x}}{x^{2}+y^{2}}\right)\!\frac{1}{2}\theta_{ij}\varepsilon^{ijab}\!=\!Z^{ab}
\end{eqnarray}
which verify
\begin{eqnarray}
X^{2}\!-\!Y^{2}\!+\!\frac{1}{8}Z^{ab}Z_{ab}\!=\!1\\
2XY\!-\!\frac{1}{16}Z^{ij}Z^{ab}\varepsilon_{ijab}\!=\!0
\end{eqnarray}
as general identities. With these one can prove that
\begin{eqnarray}
\boldsymbol{\Lambda}\!=\!
X\mathbb{I}\!+\!Yi\boldsymbol{\pi}+\frac{1}{2}Z^{ab}\boldsymbol{\sigma}_{ab}
\end{eqnarray}
in the most compact way. The inverse is given by
\begin{eqnarray}
&\boldsymbol{\Lambda}^{-1}\!=\!e^{-\frac{1}{2}\theta_{ab}\boldsymbol{\sigma}^{ab}}
\end{eqnarray}
and it can be written explicitly like
\begin{eqnarray}
\boldsymbol{\Lambda}^{-1}\!=\!X\mathbb{I}\!+\!Yi\boldsymbol{\pi}-\frac{1}{2}Z^{ab}\boldsymbol{\sigma}_{ab}
\end{eqnarray}
as clear after using relations $8X^{2}\!-\!8Y^{2}\!+\!Z^{ab}Z_{ab}\!=\!8$ and $32XY\!-\!Z^{ij}Z^{ab}\varepsilon_{ijab}\!=\!0$ as they are given here above. The complete Lorentz and phase transformation is therefore
\begin{eqnarray}
&\boldsymbol{S}\!=\!\boldsymbol{\Lambda}e^{iq\alpha}\!=\!
(X\mathbb{I}\!+\!Yi\boldsymbol{\pi}\!+\!\frac{1}{2}Z^{ab}\boldsymbol{\sigma}_{ab})e^{iq\alpha}
\end{eqnarray}
and it is called spinorial transformation. Notice that
\begin{eqnarray}
(\Lambda)^{a}_{\phantom{a}b}\boldsymbol{S}\boldsymbol{\gamma}^{b}\boldsymbol{S}^{-1}\!=\!\boldsymbol{\gamma}^{a}\label{constgamma}
\end{eqnarray}
where $(\Lambda)^{a}_{\phantom{a}b}$ such that $(\Lambda)^{a}_{\phantom{a}c}(\Lambda)^{b}_{\phantom{b}d}\eta^{cd}\!=\!\eta^{ab}$ belongs to the $\mathrm{SO(1,3)}$ group as the real Lorentz transformation \cite{Fabbri:2021mfc}.

The spinorial fields are defined as what transforms like
\begin{eqnarray}
&\psi\!\rightarrow\!\boldsymbol{S}\psi\ \ \ \ \ \ \ \ \mathrm{and} \ \ \ \ \ \ \ \ 
\overline{\psi}\!\rightarrow\!\overline{\psi}\boldsymbol{S}^{-1}
\end{eqnarray}
where $\overline{\psi}\!=\!\psi^{\dagger}\boldsymbol{\gamma}^{0}$ in the most general circumstance.

We will also introduce the object
\begin{eqnarray}
\nonumber
&(\partial_{\mu}XZ^{ab}-X\partial_{\mu}Z^{ab})
+\frac{1}{2}(\partial_{\mu}YZ_{ij}-Y\partial_{\mu}Z_{ij})\varepsilon^{ijab}+\\
&+\partial_{\mu}Z^{ak}Z^{b}_{\phantom{b}k}\!=\!-\partial_{\mu}\zeta^{ab}
\end{eqnarray}
with which we can write
\begin{eqnarray}
\boldsymbol{S}^{-1}\partial_{\mu}\boldsymbol{S}
\!=\!\frac{1}{2}\partial_{\mu}\zeta_{ab}\boldsymbol{\sigma}^{ab}\!+\!iq\partial_{\mu}\alpha
\end{eqnarray}
holding in general, and which will be useful next \cite{Fabbri:2021mfc}.

The differential character will be introduced restricted to verify the conditions 
$\boldsymbol{\nabla}_{\nu}\boldsymbol{\gamma}^{a}\!=\!0$ resulting into
\begin{eqnarray}
&\boldsymbol{\Omega}_{\mu}\!=\!\frac{1}{2}\Omega_{ij\mu}\boldsymbol{\sigma}^{ij}
\!+\!iqA_{\mu}\boldsymbol{\mathbb{I}}\label{spinorialconnection}
\end{eqnarray}
as the decomposition of the spinorial connection. 

We eventually have that
\begin{eqnarray}
&\boldsymbol{\nabla}_{\mu}\psi\!=\!\partial_{\mu}\psi
\!+\!\frac{1}{2}\Omega_{ij\mu}\boldsymbol{\sigma}^{ij}\psi\!+\!iqA_{\mu}\psi\label{spincovder}
\end{eqnarray}
is the spinorial covariant derivative of the spinor fields.

The Maxwell strength and Riemann curvature are
\begin{eqnarray}
&R^{i}_{\phantom{i}j\mu\nu}\!=\!\partial_{\mu}\Omega^{i}_{\phantom{i}j\nu}
\!-\!\partial_{\nu}\Omega^{i}_{\phantom{i}j\mu}
\!\!+\!\Omega^{i}_{\phantom{i}k\mu}\Omega^{k}_{\phantom{k}j\nu}
\!-\!\Omega^{i}_{\phantom{i}k\nu}\Omega^{k}_{\phantom{k}j\mu}\\
&F_{\mu\nu}\!=\!\partial_{\mu}A_{\nu}\!-\!\partial_{\nu}A_{\mu}
\end{eqnarray}
as consequence of commutators of covariant derivatives.

Indeed we have that we can write
\begin{eqnarray}
&[\boldsymbol{\nabla}_{\mu},\!\boldsymbol{\nabla}_{\nu}]\psi
\!=\!\frac{1}{2}R_{ij\mu\nu}\boldsymbol{\sigma}^{ij}\psi\!+\!iqF_{\mu\nu}\psi
\end{eqnarray}
in which the absence of first-order derivative indicates an absence of torsion in the covariant derivatives. Therefore this differential structure is not the most general of cases.

However, generality is restored in the dynamics by considering the Dirac matter field equations given by
\begin{eqnarray}
&i\boldsymbol{\gamma}^{\mu}\boldsymbol{\nabla}_{\mu}\psi
\!-\!XW_{\mu}\boldsymbol{\gamma}^{\mu}\boldsymbol{\pi}\psi\!-\!m\psi\!=\!0
\label{Dirac}
\end{eqnarray}
in which the $W_{\mu}$ is the axial-vector dual of the completely antisymmetric torsion and with $X$ being the torsion-spin coupling constant. This is the most general case of all.
\section{The Tensorial Connection}
\label{B}
We now write the general theory of spinor fields in its polar form, where the tensorial connection can eventually be defined. For this, we give the following theorem.

Given a spinor field, if $i\overline{\psi}\boldsymbol{\pi}\psi$ and $\overline{\psi}\psi$ are not both equal to zero identically, we can always find a frame in which
\begin{eqnarray}
&\!\psi\!=\!\phi e^{-\frac{i}{2}\beta\boldsymbol{\pi}}
\boldsymbol{L}^{-1}\left(\!\begin{tabular}{c}
$1$\\
$0$\\
$1$\\
$0$
\end{tabular}\!\right)
\label{regular}
\end{eqnarray}
for some Lorentz transformation $\boldsymbol{L}$ and in which $\phi$ and $\beta$ are a scalar and a pseudo-scalar, and the only $2$ true degrees of freedom, called module and chiral angle. Then
\begin{eqnarray}
&i\overline{\psi}\boldsymbol{\pi}\psi\!=\!2\phi^{2}\sin{\beta}\\
&\overline{\psi}\psi\!=\!2\phi^{2}\cos{\beta}
\end{eqnarray}
as well as 
\begin{eqnarray}
&\overline{\psi}\boldsymbol{\gamma}^{a}\boldsymbol{\pi}\psi\!=\!2\phi^{2}s^{a}\\
&\overline{\psi}\boldsymbol{\gamma}^{a}\psi\!=\!2\phi^{2}u^{a}
\end{eqnarray}
in general. If instead it is $i\overline{\psi}\boldsymbol{\pi}\psi\!\equiv\!\overline{\psi}\psi\!\equiv\!0$ then it is always possible to find a frame in which the spinor is
\begin{eqnarray}
&\psi\!=\!\frac{1}{\sqrt{2}}(\mathbb{I}\cos{\frac{\alpha}{2}}
\!-\!\boldsymbol{\pi}\sin{\frac{\alpha}{2}})\boldsymbol{L}^{-1}\left(\!\begin{tabular}{c}
$1$\\
$0$\\
$0$\\
$1$
\end{tabular}\!\right)
\label{singular}
\end{eqnarray}
for some Lorentz transformation $\boldsymbol{L}$ and in which $\alpha$ is a pseudo-scalar, and the only true degree of freedom. Then
\begin{eqnarray}
&\overline{\psi}\boldsymbol{\gamma}^{k}\boldsymbol{\pi}\psi
\!=\!-\sin{\alpha}\overline{\psi}\boldsymbol{\gamma}^{k}\psi
\end{eqnarray}
and
\begin{eqnarray}
&\overline{\psi}\boldsymbol{\gamma}^{k}\psi\!=\!U^{k}\\
&2i\overline{\psi}\boldsymbol{\sigma}^{ij}\psi\!=\!M^{ij}
\end{eqnarray}
in general. In this last instance, we can separate two more sub-cases: $\alpha\!=\!\pm\pi/2$ gives left-handed and right-handed chiral spinors (for which $2i\overline{\psi}\boldsymbol{\sigma}^{ij}\psi\!\equiv\!0$), known as Weyl spinors; $\alpha\!=\!0$ results into charge-conjugated spinors (for which $\overline{\psi}\boldsymbol{\gamma}^{k}\boldsymbol{\pi}\psi\!\equiv\!0$), known as Majorana spinors \cite{Fabbri:2020ypd, Fabbri:2020elt}.

In all these cases, we can introduce the object given by
\begin{eqnarray}
\boldsymbol{L}^{-1}\partial_{\mu}\boldsymbol{L}
\!=\!\frac{1}{2}\partial_{\mu}\xi^{ab}\boldsymbol{\sigma}_{ab}\!+\!iq\partial_{\mu}\xi\mathbb{I}
\label{LdL}
\end{eqnarray}
where $\xi^{ab}$ and $\xi$ are Goldstone states. Then we can define
\begin{eqnarray}
&\partial_{\mu}\xi_{ij}\!-\!\Omega_{ij\mu}\!\equiv\!R_{ij\mu}\label{R}\\
&q(\partial_{\mu}\xi\!-\!A_{\mu})\!\equiv\!P_{\mu}\label{P}
\end{eqnarray}
which can be demonstrated to be real tensors, although they contain the same amount of information as spin connection and gauge potential, and therefore they are called tensorial connection and gauge momentum \cite{Fabbri:2018crr}.

With them the spinorial covariant derivative of (\ref{regular}) is
\begin{eqnarray}
&\!\!\!\!\!\!\!\!\boldsymbol{\nabla}_{\mu}\psi\!=\!(-\frac{i}{2}\nabla_{\mu}\beta\boldsymbol{\pi}
\!+\!\nabla_{\mu}\ln{\phi}\mathbb{I}\!-\!\frac{1}{2}R_{ij\mu}\boldsymbol{\sigma}^{ij}
\!-\!iP_{\mu}\mathbb{I})\psi\label{decspinder}
\end{eqnarray}
from which
\begin{eqnarray}
&\nabla_{\mu}u_{i}\!=\!R_{ji\mu}u^{j}\label{u}\\
&\nabla_{\mu}s_{i}\!=\!R_{ji\mu}s^{j}\label{s}
\end{eqnarray}
while that of (\ref{singular}) is
\begin{eqnarray}
\nonumber
&\boldsymbol{\nabla}_{\mu}\psi\!=\![-\frac{1}{2}(\mathbb{I}\tan{\alpha}
\!+\!\boldsymbol{\pi}\sec{\alpha})\nabla_{\mu}\alpha-\\
&-\frac{1}{2}R_{ij\mu}\boldsymbol{\sigma}^{ij}\!-\!iP_{\mu}\mathbb{I}]\psi
\end{eqnarray}
from which
\begin{eqnarray}
&\!\!\nabla_{\mu}U_{i}\!=\!R_{ji\mu}U^{j}\label{U}\\
&\!\!\!\!\nabla_{\mu}M^{ab}\!=\!-M^{ab}\tan{\alpha}\nabla_{\mu}\alpha
\!-\!R^{a}_{\phantom{a}k\mu}M^{kb}\!+\!R^{b}_{\phantom{b}k\mu}M^{ka}\label{M}
\end{eqnarray}
written in terms of the derivatives of the degrees of freedom plus the contributions of the tensorial connection as well as the momentum. However, for Weyl spinors
\begin{eqnarray}
&\boldsymbol{\nabla}_{\mu}\psi
\!=\!(-\frac{1}{2}R_{ij\mu}\boldsymbol{\sigma}^{ij}\!-\!iP_{\mu}\mathbb{I})\psi
\label{w}
\end{eqnarray}
only with tensorial connection and momentum while for the Majorana spinors we have that
\begin{eqnarray}
&\boldsymbol{\nabla}_{\mu}\psi\!=\!-\frac{1}{2}R_{ij\mu}\boldsymbol{\sigma}^{ij}\psi
\label{m}
\end{eqnarray}
only with the tensorial connection \cite{Fabbri:2020ypd, Fabbri:2020elt}.

The Maxwell strength and the Riemann curvature are
\begin{eqnarray}
&\!\!\!\!\!\!\!\!R^{i}_{\phantom{i}j\mu\nu}\!=\!-(\nabla_{\mu}R^{i}_{\phantom{i}j\nu}
\!-\!\!\nabla_{\nu}R^{i}_{\phantom{i}j\mu}
\!\!+\!R^{i}_{\phantom{i}k\mu}R^{k}_{\phantom{k}j\nu}
\!-\!R^{i}_{\phantom{i}k\nu}R^{k}_{\phantom{k}j\mu})\label{Riemann}\\
&\!\!\!\!\!\!\!\!qF_{\mu\nu}\!=\!-(\nabla_{\mu}P_{\nu}
\!-\!\!\nabla_{\nu}P_{\mu})\label{Maxwell}
\end{eqnarray}
as it can be checked straightforwardly and as usual they contain information about electrodynamics and gravitation alone. Therefore if it were possible to find non-zero $R_{ij\mu}$ and $P_{\mu}$ such that $R_{ij\mu\nu}\!\equiv\!0$ and $F_{\mu\nu}\!\equiv\!0$ they would describe tensorial connection and gauge momentum that are non-trivial but which are unrelated to any source.\! For such a situation, we have that the tensorial connection is encoding information about frames only while the gauge momentum is encoding information about gauges solely.

The Dirac equations for (\ref{regular}) become
\begin{eqnarray}
&\!\!\!\!B_{\mu}\!-\!2P^{\iota}u_{[\iota}s_{\mu]}\!+\!(\nabla\beta\!-\!2XW)_{\mu}
\!+\!2s_{\mu}m\cos{\beta}\!=\!0\label{dep1}\\
&\!\!\!\!R_{\mu}\!-\!2P^{\rho}u^{\nu}s^{\alpha}\varepsilon_{\mu\rho\nu\alpha}\!+\!2s_{\mu}m\sin{\beta}
\!+\!\nabla_{\mu}\ln{\phi^{2}}\!=\!0\label{dep2}
\end{eqnarray}
with $R_{\mu a}^{\phantom{\mu a}a}\!=\!R_{\mu}$ and $\frac{1}{2}\varepsilon_{\mu\alpha\nu\iota}R^{\alpha\nu\iota}\!=\!B_{\mu}$ specify all derivatives of module and chiral angle while for (\ref{singular}) we have
\begin{eqnarray}
\nonumber
&\![(2XW\!-\!B)^{\sigma}\varepsilon_{\sigma\mu\rho\nu}\!+\!R_{[\mu}g_{\rho]\nu}+\\
&+g_{\nu[\mu}\nabla_{\rho]}\alpha\tan{\alpha}]
M_{\eta\zeta}\varepsilon^{\mu\rho\eta\zeta}\!=\!0\label{f1}\\
\nonumber
&\!\![(2XW\!-\!B)^{\sigma}\varepsilon_{\sigma\mu\rho\nu}\!+\!R_{[\mu}g_{\rho]\nu}+\\
&+g_{\nu[\mu}\nabla_{\rho]}\alpha\tan{\alpha}]M^{\mu\rho}\!+\!4mU_{\nu}\!=\!0\label{f2}\\
&\!\!\!\!(\varepsilon^{\mu\rho\sigma\nu}\nabla_{\mu}\alpha\sec{\alpha}
\!-\!2P^{[\rho}g^{\sigma]\nu})M_{\rho\sigma}\!=\!0\label{f3}\\
&\!\!\!\!\!\!\!\!M_{\rho\sigma}(g^{\nu[\rho}\nabla^{\sigma]}\alpha\sec{\alpha}
\!-\!2P_{\mu}\varepsilon^{\mu\rho\sigma\nu})\!+\!4m\sin{\alpha}U^{\nu}\!=\!0\label{f4}
\end{eqnarray}
specifying all derivatives for the only degree of freedom they have. In particular, for Weyl spinors $m\!=\!0$ and
\begin{eqnarray}
&R_{\mu}U^{\mu}\!=\!0\label{w1}\\
&(-B_{\mu}\!+\!2XW_{\mu}\!\pm\!2P_{\mu})U^{\mu}\!=\!0\label{w2}\\
&[(-B_{\mu}\!+\!2XW_{\mu}\!\pm\!2P_{\mu})\varepsilon^{\mu\rho\alpha\nu}
\!+\!g^{\rho[\alpha}R^{\nu]}]U_{\rho}\!=\!0\label{w3}
\end{eqnarray}
while for Majorana spinors
\begin{eqnarray}
&(g_{\sigma[\pi}B_{\kappa]}\!-\!R^{\mu}\varepsilon_{\mu\sigma\pi\kappa})M^{\pi\kappa}
\!=\!0\label{m1}\\
&\frac{1}{2}(B_{\mu}\varepsilon^{\mu\sigma\pi\kappa}
\!+\!g^{\sigma[\pi}R^{\kappa]})M_{\pi\kappa}\!-\!2mU^{\sigma}\!=\!0\label{m2}
\end{eqnarray}
with no torsion since Majorana spinors have identically zero spin axial-vector and torsion decouples. Notice that all these are equivalent to the Dirac equations \cite{Fabbri:2020ypd, Fabbri:2020elt}.
\section{Special Backgrounds}
\label{C}
In section \ref{A}-\ref{B} we introduced the tools leading to the definition of tensorial connections. In particular, toward the end of section \ref{B} we have also stated that it is possible to find non-zero tensorial connections that have zero curvature tensor. Now we shall provide two examples.

Both will be given in spherical coordinates with metric
\begin{eqnarray}
&g_{tt}\!=\!1\\
&g_{rr}\!=\!-1\\
&g_{\theta\theta}\!=\!-r^{2}\\
&g_{\varphi\varphi}\!=\!-r^{2}|\!\sin{\theta}|^{2}
\end{eqnarray}
and connection
\begin{eqnarray}
&\Lambda^{\theta}_{\theta r}\!=\!\frac{1}{r}\\
&\Lambda^{\varphi}_{\varphi r}\!=\!\frac{1}{r}\\
&\Lambda^{r}_{\theta\theta}\!=\!-r\\
&\Lambda^{r}_{\varphi\varphi}\!=\!-r|\!\sin{\theta}|^{2}\\
&\Lambda^{\varphi}_{\varphi\theta}\!=\!\cot{\theta}\\
&\Lambda^{\theta}_{\varphi\varphi}\!=\!-\cos{\theta}\sin{\theta}
\end{eqnarray}
having zero curvature as is well known.

The \emph{first example} is already well known as it has been established in reference \cite{Fabbri:2019kfr}. It is given for tetrads
\begin{eqnarray}
&\!\!\!\!\xi^{0}_{t}\!=\!\cosh{\alpha}\ \ \ \ 
\xi^{2}_{t}\!=\!\sinh{\alpha}\\
&\!\!\!\!\xi^{1}_{r}\!=\!\sin{\gamma}\ \ \ \ 
\xi^{3}_{r}\!=\!-\cos{\gamma}\\
&\!\!\!\!\xi^{1}_{\theta}\!=\!-r\cos{\gamma}\ \ \ \ 
\xi^{3}_{\theta}\!=\!-r\sin{\gamma}\\
&\!\!\!\!\xi^{0}_{\varphi}\!=\!r\sin{\theta}\sinh{\alpha}\ \ \ \ 
\xi^{2}_{\varphi}\!=\!r\sin{\theta}\cosh{\alpha}
\end{eqnarray}
and dual tetrads
\begin{eqnarray}
&\!\!\!\!\xi_{0}^{t}\!=\!\cosh{\alpha}\ \ \ \ \xi_{2}^{t}\!=\!-\sinh{\alpha}\\
&\!\!\!\!\xi_{1}^{r}\!=\!\sin{\gamma}\ \ \ \ \xi_{3}^{r}\!=\!-\cos{\gamma}\\
&\!\!\!\!\xi_{1}^{\theta}\!=\!-\frac{1}{r}\cos{\gamma}\ \ \ \ 
\xi_{3}^{\theta}\!=\!-\frac{1}{r}\sin{\gamma}\\
&\!\!\!\!\xi_{0}^{\varphi}\!=\!-\frac{1}{r\sin{\theta}}\sinh{\alpha}\ \ \ \ 
\xi_{2}^{\varphi}\!=\!\frac{1}{r\sin{\theta}}\cosh{\alpha}
\end{eqnarray}
from which the spin connection is
\begin{eqnarray}
&\Omega_{02r}\!=\!-\partial_{r}\alpha
\ \ \ \ \Omega_{13r}\!=\!-\partial_{r}(\theta\!+\!\gamma)\\
&\Omega_{02\theta}\!=\!-\partial_{\theta}\alpha
\ \ \ \ \Omega_{13\theta}\!=\!-\partial_{\theta}(\theta\!+\!\gamma)\\
&\Omega_{01\varphi}\!=\!-\cos{(\theta\!+\!\gamma)}\sinh{\alpha}\\ 
&\Omega_{03\varphi}\!=\!-\sin{(\theta\!+\!\gamma)}\sinh{\alpha}\\
&\Omega_{23\varphi}\!=\!\sin{(\theta\!+\!\gamma)}\cosh{\alpha}\\ 
&\Omega_{12\varphi}\!=\!-\cos{(\theta\!+\!\gamma)}\cosh{\alpha}
\end{eqnarray}
in which $\alpha\!=\!\alpha(r,\theta)$ and $\gamma\!=\!\gamma(r,\theta)$ are generic functions and
for which the ensuing curvature is vanishing.

Our first example is built to be compatible with spinors of the type (\ref{regular}). In this case we have that
\begin{eqnarray}
&u_{t}\!=\!\cosh{\alpha}\label{u1}\ \ \ \ \ \ \ \
&u_{\varphi}\!=\!r\sin{\theta}\sinh{\alpha}\label{u2}
\end{eqnarray}
\begin{eqnarray}
&s_{r}\!=\!\cos{\gamma}\label{s1}\ \ \ \ \ \ \ \
&s_{\theta}\!=\!r\sin{\gamma}\label{s2}
\end{eqnarray}
and because of (\ref{u}-\ref{s}) we can see that
\begin{eqnarray}
&r\sin{\theta}\partial_{\theta}\alpha\!=\!R_{t\varphi\theta}\label{dA1}\\
&r\sin{\theta}\partial_{r}\alpha\!=\!R_{t\varphi r}\label{dA2}\\
&-r(1\!+\!\partial_{\theta}\gamma)\!=\!R_{r\theta\theta}\label{dA3}\\
&r\partial_{r}\gamma\!=\!R_{\theta rr}\label{dA4}
\end{eqnarray}
\begin{eqnarray}
&\!rR_{rt\varphi}\!=\!R_{t\theta\varphi}\tan{\gamma}\label{aux1}\\
&\!\!r\sin{\theta}R_{t\theta\varphi}\!=\!(R_{\varphi\theta\varphi}
\!-\!r^{2}\cos{\theta}\sin{\theta})\tanh{\alpha}\label{aux2}\\
&\!\!\!\!(R_{\varphi\theta\varphi}\!-\!r^{2}\sin{\theta}\cos{\theta})\tan{\gamma}\!=\!
r(R_{r\varphi\varphi}\!+\!r|\!\sin{\theta}|^{2})\label{aux3}\\
&\!\!(R_{r\varphi\varphi}\!+\!r|\!\sin{\theta}|^{2})\tanh{\alpha}
\!=\!r\sin{\theta}R_{rt\varphi}\label{aux4}
\end{eqnarray}
\begin{eqnarray}
&rR_{rtt}\!=\!R_{t\theta t}\tan{\gamma}\\
&r\sin{\theta}R_{t\theta t}\!=\!R_{\varphi\theta t}\tanh{\alpha}\\
&R_{\varphi\theta t}\tan{\gamma}\!=\!rR_{r\varphi t}\\
&R_{r\varphi t}\tanh{\alpha}\!=\!r\sin{\theta}R_{rtt}
\end{eqnarray}
\begin{eqnarray}
&rR_{rtr}\!=\!R_{t\theta r}\tan{\gamma}\\
&r\sin{\theta}R_{t\theta r}\!=\!R_{\varphi\theta r}\tanh{\alpha}\\
&R_{\varphi\theta r}\tan{\gamma}\!=\!rR_{r\varphi r}\\
&R_{r\varphi r}\tanh{\alpha}\!=\!r\sin{\theta}R_{rtr}
\end{eqnarray}
\begin{eqnarray}
&rR_{rt\theta}\!=\!R_{t\theta\theta}\tan{\gamma}\\
&r\sin{\theta}R_{t\theta\theta}\!=\!R_{\varphi\theta\theta}\tanh{\alpha}\\
&R_{\varphi\theta\theta}\tan{\gamma}\!=\!rR_{r\varphi\theta}\\
&R_{r\varphi\theta}\tanh{\alpha}\!=\!r\sin{\theta}R_{rt\theta}
\end{eqnarray}
grouped in four blocks of four relations. So as educated guess we can choose (\ref{dA1}-\ref{dA4}) accompanied by the
\begin{eqnarray}
&R_{r\varphi\varphi}\!=\!-r|\!\sin{\theta}|^{2}\\
&R_{\theta\varphi\varphi}\!=\!-r^{2}\cos{\theta}\sin{\theta}
\end{eqnarray}
\begin{eqnarray}
&R_{rtt}\!=\!2\varepsilon\sinh{\alpha}\sin{\gamma}\\
&R_{\varphi rt}\!=\!-2\varepsilon r\sin{\theta}\cosh{\alpha}\sin{\gamma}\\
&R_{\theta tt}\!=\!-2\varepsilon r\sinh{\alpha}\cos{\gamma}\\
&R_{\varphi\theta t}\!=\!2\varepsilon r^{2}\sin{\theta}\cosh{\alpha}\cos{\gamma}
\end{eqnarray}
with $\varepsilon$ being a generic constant because of the vanishing of the curvature tensor, that is an integration constant.

In this first example, we have that
\begin{eqnarray}
&R_{r}\!=\!2\varepsilon\sinh{\alpha}\sin{\gamma}\!+\!(2\!+\!\partial_{\theta}\gamma)/r\\
&R_{\theta}\!=\!-2\varepsilon r \sinh{\alpha}\cos{\gamma}\!-\!r\partial_{r}\gamma\!+\!\cot{\theta}
\end{eqnarray}
\begin{eqnarray}
&B_{r}\!=\!2\varepsilon\cosh{\alpha}\cos{\gamma}\!+\!\frac{1}{r}\partial_{\theta}\alpha\\
&B_{\theta}\!=\!2\varepsilon r\cosh{\alpha}\sin{\gamma}\!-\!r\partial_{r}\alpha
\end{eqnarray}
as the components of the tensorial connection that enter into the field equations. The field equations (\ref{dep1}-\ref{dep2}) with no torsion and for $P_{t}\!=\!m$ are therefore given by
\begin{eqnarray}
\nonumber
&\!\!\!\!\partial_{\theta}\alpha\!-\!2(m\!-\!\varepsilon)r\cosh{\alpha}\cos{\gamma}+\\
&+r\partial_{r}\beta\!+\!2mr\cos{\beta}\cos{\gamma}\!=\!0
\end{eqnarray}
\begin{eqnarray}
\nonumber
&\!\!\!\!r\partial_{r}\alpha\!+\!2(m\!-\!\varepsilon)r\cosh{\alpha}\sin{\gamma}-\\
&-\partial_{\theta}\beta\!-\!2mr\cos{\beta}\sin{\gamma}\!=\!0
\end{eqnarray}
\begin{eqnarray}
\nonumber
&\!\!\partial_{\theta}\gamma\!-\!2(m\!-\!\varepsilon)r\sin{\gamma}\sinh{\alpha}+\\
&+2mr\cos{\gamma}\sin{\beta}
\!+\!r\partial_{r}\ln{(\phi^{2}r^{2}\sin{\theta})}\!=\!0
\end{eqnarray}
\begin{eqnarray}
\nonumber
&\!\!-r\partial_{r}\gamma\!+\!2(m\!-\!\varepsilon)r\cos{\gamma}\sinh{\alpha}+\\
&+2mr\sin{\gamma}\sin{\beta}
\!+\!\partial_{\theta}\ln{(\phi^{2}r^{2}\sin{\theta})}\!=\!0
\end{eqnarray}
in the most general case.

Looking for solutions with no torsion we assume $\beta\!=\!0$ for consistency. In this case the field equations become
\begin{eqnarray}
&\partial_{\theta}\alpha\!-\!2(m\!-\!\varepsilon)r\cosh{\alpha}\cos{\gamma}
\!+\!2mr\cos{\gamma}\!=\!0\label{deps1}\\
&\partial_{r}\alpha\!+\!2(m\!-\!\varepsilon)\cosh{\alpha}\sin{\gamma}
\!-\!2m\sin{\gamma}\!=\!0\label{deps2}
\end{eqnarray}
\begin{eqnarray}
&\partial_{\theta}\gamma\!-\!2(m\!-\!\varepsilon)r\sin{\gamma}\sinh{\alpha}
\!+\!r\partial_{r}F\!=\!0\label{deps3}\\
&-r\partial_{r}\gamma\!+\!2(m\!-\!\varepsilon)r\cos{\gamma}\sinh{\alpha}
\!+\!\partial_{\theta}F\!=\!0\label{deps4}
\end{eqnarray}
where $\ln(\phi^{2}r^{2}\sin{\theta})\!=\!F$ was set. Alternatively
\begin{eqnarray}
&\partial_{\theta}\alpha\!=\![2(m\!-\!\varepsilon)\cosh{\alpha}\!-\!2m]r\cos{\gamma}\\
&\partial_{r}\alpha\!=\!-[2(m\!-\!\varepsilon)\cosh{\alpha}\!-\!2m]\sin{\gamma}
\end{eqnarray}
\begin{eqnarray}
&\partial_{\theta}\gamma\!=\!-r\partial_{r}A\\
&r\partial_{r}\gamma\!=\!\partial_{\theta}A
\end{eqnarray}
having set $\ln|2(m\!-\!\varepsilon)\cosh{\alpha}\!-\!2m|\!+\!F\!=\!A$ so to make the expressions symmetric. Hence imposing the integrability conditions $\partial_{\theta}\partial_{r}\alpha\!=\!\partial_{r}\partial_{\theta}\alpha$ and $\partial_{\theta}\partial_{r}A\!=\!\partial_{r}\partial_{\theta}A$ we get
\begin{eqnarray}
\cos{\gamma}(1\!+\!\partial_{\theta}\gamma)\!=\!\sin{\gamma}r\partial_{r}\gamma\label{constraint}
\end{eqnarray}
\begin{eqnarray}
&\partial_{\theta}\partial_{\theta}\gamma\!+\!r\partial_{r}(r\partial_{r}\gamma)\!=\!0
\end{eqnarray}
showing that $\gamma$ is harmonic. The only harmonic functions that satisfy (\ref{constraint}) are $\sin{\gamma}\!=\!\pm1$ and $\gamma\!=\!-\theta$ in general.

In the above derivation we assumed $\alpha$ variable, which is generally true, although we can still assume $\alpha$ constant and such that its value is fixed to
\begin{eqnarray}
&\cosh{\alpha}\!=\!m/(m\!-\!\varepsilon)
\end{eqnarray}
as solution of (\ref{deps1}-\ref{deps2}). In this case (\ref{deps3}-\ref{deps4}) become
\begin{eqnarray}
&\partial_{\theta}\gamma\!-\![2(m\!-\!\varepsilon)\sinh{\alpha}]r\sin{\gamma}
\!+\!r\partial_{r}F\!=\!0\\
&-r\partial_{r}\gamma\!+\![2(m\!-\!\varepsilon)\sinh{\alpha}]r\cos{\gamma}
\!+\!\partial_{\theta}F\!=\!0
\end{eqnarray}
to be worked out. Multiplying with $e^{F}\cos{\gamma}$ and $e^{F}\sin{\gamma}$ we get the set of conditions
\begin{eqnarray}
&\!\!\partial_{\theta}(e^{F}\!\sin{\gamma})\!+\!r\partial_{r}(e^{F}\!\cos{\gamma})\!=\!0\\
&\!\!\!\!r\partial_{r}(e^{F}\!\sin{\gamma})\!-\!\partial_{\theta}(e^{F}\!\cos{\gamma})
\!=\![2(m\!-\!\varepsilon)\sinh{\alpha}]re^{F}
\end{eqnarray}
the first one of which being integrated as 
\begin{eqnarray}
&e^{F}\!\sin{\gamma}\!=\!r\partial_{r}G\\
&e^{F}\!\cos{\gamma}\!=\!-\partial_{\theta}G
\end{eqnarray}
so that
\begin{eqnarray}
\nonumber
&\!\!r\partial_{r}(r\partial_{r}G)\!+\!\partial_{\theta}\partial_{\theta}G=\\
&=[2(m\!-\!\varepsilon)\sinh{\alpha}]r\sqrt{|r\partial_{r}G|^{2}\!+\!|\partial_{\theta}G|^{2}}
\end{eqnarray}
as a Laplace equation with a very specific source. Due to its non-linearity, we cannot hope to find solutions, albeit it is possible to check that $\sin{\gamma}\!=\!\pm1$ and $\gamma\!=\!-\theta$ still are.

Picking $\sin{\gamma}\!=\!-1$ we have that
\begin{eqnarray}
&\partial_{\theta}\alpha\!+\!r\partial_{r}\beta\!=\!0\label{deps1a}\\
&r\partial_{r}\alpha\!-\!2(m\!-\!\varepsilon)r\cosh{\alpha}
\!-\!\partial_{\theta}\beta\!+\!2mr\cos{\beta}\!=\!0\label{deps2a}\\
&2(m\!-\!\varepsilon)\sinh{\alpha}
\!+\!\partial_{r}\ln{(\phi^{2}r^{2}\sin{\theta})}\!=\!0\label{deps3a}\\
&-2mr\sin{\beta}\!+\!\partial_{\theta}\ln{(\phi^{2}r^{2}\sin{\theta})}\!=\!0\label{deps4a}
\end{eqnarray}
allowing to take $\sinh{\alpha}\!=\!\sqrt{\varepsilon(2m\!-\!\varepsilon)}/(m\!-\!\varepsilon)$ giving
\begin{eqnarray}
&\beta\!=\!0\\
&\phi^{2}r^{2}\sin{\theta}\!=\!e^{-2r\sqrt{\varepsilon(2m-\varepsilon)}}\label{1}
\end{eqnarray}
as an exact solution.\! Taking $m\!>\!\varepsilon\!>\!0$ this \emph{first solution} is square-integrable (although its energy has a logarithmic divergence in the origin of the coordinates).

The other interesting case is when $\gamma\!=\!-\theta$ so that
\begin{eqnarray}
\nonumber
&\!\!\!\!\partial_{\theta}\alpha\!-\!2(m\!-\!\varepsilon)r\cosh{\alpha}\cos{\theta}+\\
&+r\partial_{r}\beta\!+\!2mr\cos{\beta}\cos{\theta}\!=\!0\label{deps1b}\\
\nonumber
&\!\!\!\!r\partial_{r}\alpha\!-\!2(m\!-\!\varepsilon)r\cosh{\alpha}\sin{\theta}-\\
&-\partial_{\theta}\beta\!+\!2mr\cos{\beta}\sin{\theta}\!=\!0\label{deps2b}\\
\nonumber
&-1\!+\!2(m\!-\!\varepsilon)r\sin{\theta}\sinh{\alpha}+\\
&+2mr\cos{\theta}\sin{\beta}
\!+\!r\partial_{r}\ln{(\phi^{2}r^{2}\sin{\theta})}\!=\!0\label{deps3b}\\
\nonumber
&2(m\!-\!\varepsilon)r\cos{\theta}\sinh{\alpha}-\\
&-2mr\sin{\theta}\sin{\beta}
\!+\!\partial_{\theta}\ln{(\phi^{2}r^{2}\sin{\theta})}\!=\!0\label{deps4b}
\end{eqnarray}
with $\sinh{\alpha}\!=\!\sqrt{\varepsilon(2m\!-\!\varepsilon)}/(m\!-\!\varepsilon)$ giving
\begin{eqnarray}
&\beta\!=\!0\\
&\phi^{2}r^{2}\sin{\theta}\!=\!re^{-2r\sin{\theta}\sqrt{\varepsilon(2m-\varepsilon)}}\label{2}
\end{eqnarray}
as exact solution.\! Such a \emph{second solution} is regular at the origin (although not of course square-integrable).

If we do not care about square-integrability, a vast class of solutions becomes possible. In (\ref{deps1a}-\ref{deps4a}) we could look for solutions where $\alpha$ is not constant but given by
\begin{eqnarray}
&\tanh{\frac{\alpha}{2}}\!=\!-\sqrt{\frac{\varepsilon}{2m-\varepsilon}}
\tanh{\left[r\sqrt{\varepsilon(2m\!-\!\varepsilon)}\right]}
\end{eqnarray}
so that
\begin{eqnarray}
&\!\!\beta\!=\!0\\
&\!\!\!\!\!\!\!\!\phi^{2}r^{2}\sin{\theta}\!=\!m
\!+\!(m\!-\!\varepsilon)\cosh{\left(2r\sqrt{\varepsilon(2m\!-\!\varepsilon)}\right)}\label{3}
\end{eqnarray}
is the form of the \emph{third solution}. It is regular in the origin of coordinates (but obviously dramatically divergent).

Instead in (\ref{deps1b}-\ref{deps4b}) we might have
\begin{eqnarray}
&\!\!\!\!\tanh{\frac{\alpha}{2}}\!=\!-\sqrt{\frac{\varepsilon}{2m-\varepsilon}}
\tanh{\left[r\sin{\theta}\sqrt{\varepsilon(2m\!-\!\varepsilon)}\right]}
\end{eqnarray}
so that
\begin{eqnarray}
&\!\!\beta\!=\!0\\
&\!\!\!\!\!\!\!\!\!\!\!\!\phi^{2}r^{2}\sin{\theta}\!=\!r\!\left[m\!+\!(m\!-\!\varepsilon)
\!\cosh{\!\left(\!2r\sin{\theta}\sqrt{\varepsilon(2m\!-\!\varepsilon)}\right)\!}\right]\label{4}
\end{eqnarray}
is the \emph{fourth solution}. As usual this is regular in the origin of the coordinates (but of course it is largely divergent).

It is worth remarking that for the $\phi^{2}r^{2}\sin{\theta}$ functions, compared to the $\sin{\gamma}\!=\!-1$ cases given by (\ref{1}, \ref{3}), the $\gamma\!=\!-\theta$ cases (\ref{2}, \ref{4}) acquire an extra $r$ dependence in the overall function and an extra $\sin{\theta}$ dependence inside the exponentials. The $r$ dependence is what makes these modules more convergent at the origin whereas the $\sin{\theta}$ dependence is what spoils the convergent behaviour they have at the infinity. Remark also that the two cases with $\alpha$ constant are in no way a limit of the corresponding two cases in which $\alpha$ is variable. So these four cases are four independent exact solutions for the same field equations.

The \emph{second example} is new to this paper.\! It is given in terms of the basis of tetrad fields
\begin{eqnarray}
&\!\!\!\!\xi^{0}_{t}\!=\!\cosh{\alpha}\ \ \ \ 
\xi^{3}_{t}\!=\!\sinh{\alpha}\\
&\!\!\!\!\xi^{2}_{r}\!=\!\sin{\gamma}\ \ \ \ 
\xi^{1}_{r}\!=\!-\cos{\gamma}\\
&\!\!\!\!\xi^{2}_{\theta}\!=\!-r\cos{\gamma}\ \ \ \ 
\xi^{1}_{\theta}\!=\!-r\sin{\gamma}\\
&\!\!\!\!\xi^{0}_{\varphi}\!=\!r\sin{\theta}\sinh{\alpha}\ \ \ \ 
\xi^{3}_{\varphi}\!=\!r\sin{\theta}\cosh{\alpha}
\end{eqnarray}
with dual tetrad fields
\begin{eqnarray}
&\!\!\!\!\xi_{0}^{t}\!=\!\cosh{\alpha}\ \ \ \ \ \ \ \ 
\xi_{3}^{t}\!=\!-\sinh{\alpha}\\
&\!\!\!\!\xi_{2}^{r}\!=\!\sin{\gamma}\ \ \ \ \ \ \ \ 
\xi_{1}^{r}\!=\!-\cos{\gamma}\\
&\!\!\!\!\xi_{2}^{\theta}\!=\!-\frac{1}{r}\cos{\gamma}\ \ \ \ 
\xi_{1}^{\theta}\!=\!-\frac{1}{r}\sin{\gamma}\\
&\!\!\!\!\xi_{0}^{\varphi}\!=\!-\frac{1}{r\sin{\theta}}\sinh{\alpha}\ \ \ \ 
\xi_{3}^{\varphi}\!=\!\frac{1}{r\sin{\theta}}\cosh{\alpha}
\end{eqnarray}
giving spin connection
\begin{eqnarray}
&\Omega_{03r}\!=\!-\partial_{r}\alpha\ \ \ \ \Omega_{21r}\!=\!-\partial_{r}(\theta\!+\!\gamma)\\
&\Omega_{03\theta}\!=\!-\partial_{\theta}\alpha\ \ \ \ \Omega_{21\theta}\!=\!-\partial_{\theta}(\theta\!+\!\gamma)\\
&\Omega_{02\varphi}\!=\!-\cos{(\theta\!+\!\gamma)}\sinh{\alpha}\\ 
&\Omega_{01\varphi}\!=\!-\sin{(\theta\!+\!\gamma)}\sinh{\alpha}\\
&\Omega_{31\varphi}\!=\!\sin{(\theta\!+\!\gamma)}\cosh{\alpha}\\ 
&\Omega_{23\varphi}\!=\!-\cos{(\theta\!+\!\gamma)}\cosh{\alpha}
\end{eqnarray}
in which $\alpha\!=\!\alpha(r,\theta)$ and $\gamma\!=\!\gamma(r,\theta)$ are generic functions and
for which the ensuing curvature is again vanishing.

Our second example is constructed to be compatible with spinors of the type (\ref{singular}). In this case
\begin{eqnarray}
&U_{t}\!=\!e^{\alpha}\ \ \ \ \ \ \ \ \ \ \ \ \ \ \ \ 
U_{\varphi}\!=\!e^{\alpha}r\sin{\theta}
\end{eqnarray}
\begin{eqnarray}
&M_{tr}\!=\!-e^{\alpha}\sin{\gamma}\ \ \ \ \ \ \ \ 
M_{r\varphi}\!=\!e^{\alpha}r\sin{\theta}\sin{\gamma}\\
&M_{t\theta}\!=\!e^{\alpha}r\cos{\gamma}\ \ \ \ 
M_{\theta\varphi}\!=\!-e^{\alpha}r^{2}\sin{\theta}\cos{\gamma}
\end{eqnarray}
and then (\ref{U}-\ref{M}) are solved for
\begin{eqnarray}
&r\sin{\theta}\partial_{\theta}\alpha\!=\!R_{t\varphi\theta}\label{dB1}\\
&r\sin{\theta}\partial_{r}\alpha\!=\!R_{t\varphi r}\label{dB2}\\
&-r(1\!+\!\partial_{\theta}\gamma)\!=\!R_{r\theta\theta}\label{dB3}\\
&r\partial_{r}\gamma\!=\!R_{\theta rr}\label{dB4}
\end{eqnarray}
\begin{eqnarray}
&-r|\!\sin{\theta}|^{2}\!-\!r\sin{\theta}R_{tr\varphi}
\!=\!R_{r\varphi\varphi}\label{bux1}\\
&-r^{2}\sin{\theta}\cos{\theta}\!-\!r\sin{\theta}R_{t\theta\varphi}
\!=\!R_{\theta\varphi\varphi}\label{bux2}
\end{eqnarray}
\begin{eqnarray}
&R_{rtt}r\sin{\theta}\!=\!R_{r\varphi t}\\
&R_{\theta tt}r\sin{\theta}\!=\!R_{\theta\varphi t}
\end{eqnarray}
\begin{eqnarray}
&R_{trr}r\sin{\theta}\!=\!R_{\varphi rr}\\
&R_{t\theta r}r\sin{\theta}\!=\!R_{\varphi\theta r}
\end{eqnarray}
\begin{eqnarray}
&R_{t\theta\theta}r\sin{\theta}\!=\!R_{\varphi\theta\theta}\\
&R_{tr\theta}r\sin{\theta}\!=\!R_{\varphi r\theta}
\end{eqnarray}
as is direct to verify. A restricted case is given when the expressions (\ref{dB1}-\ref{dB4}) are accompanied by
\begin{eqnarray}
&R_{r\varphi\varphi}\!=\!-r|\!\sin{\theta}|^{2}\label{r1}\\
&R_{\theta\varphi\varphi}\!=\!-r^{2}\cos{\theta}\sin{\theta}\label{r2}
\end{eqnarray}
\begin{eqnarray}
&R_{rtt}\!=\!A\\
&R_{\theta tt}\!=\!rB\\
&R_{r\varphi t}\!=\!Ar\sin{\theta}\\
&R_{\theta\varphi t}\!=\!Br^{2}\sin{\theta}
\end{eqnarray}
with the $A\!=\!A(r,\theta)$ and $B\!=\!B(r,\theta)$ being generic functions that have to be determined by the integrability conditions arising from the vanishing of the curvature. These integrability conditions are
\begin{eqnarray}
\partial_{r}A-A\partial_{r}\alpha+B\partial_{r}\gamma\!=\!0\\
\partial_{\theta}A-A\partial_{\theta}\alpha+B\partial_{\theta}\gamma\!=\!0\\
\partial_{r}B-B\partial_{r}\alpha-A\partial_{r}\gamma\!=\!0\\
\partial_{\theta}B-B\partial_{\theta}\alpha-A\partial_{\theta}\gamma\!=\!0
\end{eqnarray}
which can be worked out, multiplying all expressions by $e^{-\alpha}\sin{\gamma}$ or $e^{-\alpha}\cos{\gamma}$ and taking linear combinations, as
\begin{eqnarray}
&\partial_{r}(Ae^{-\alpha}\sin{\gamma}\!-\!Be^{-\alpha}\cos{\gamma})\!=\!0\\
&\partial_{\theta}(Ae^{-\alpha}\sin{\gamma}\!-\!Be^{-\alpha}\cos{\gamma})\!=\!0\\
&\partial_{r}(Be^{-\alpha}\sin{\gamma}\!+\!Ae^{-\alpha}\cos{\gamma})\!=\!0\\
&\partial_{\theta}(Be^{-\alpha}\sin{\gamma}\!+\!Ae^{-\alpha}\cos{\gamma})\!=\!0
\end{eqnarray}
so that after integration
\begin{eqnarray}
&Ae^{-\alpha}\sin{\gamma}-Be^{-\alpha}\cos{\gamma}\!=\!a\\
&Be^{-\alpha}\sin{\gamma}+Ae^{-\alpha}\cos{\gamma}\!=\!b
\end{eqnarray}
and a quick diagonalization
\begin{eqnarray}
&A\!=\!e^{\alpha}(a\sin{\gamma}\!+\!b\cos{\gamma})\\
&B\!=\!e^{\alpha}(b\sin{\gamma}\!-\!a\cos{\gamma})
\end{eqnarray}
with $a$ and $b$ generic constants coming from the vanishing of the curvature tensor, and so two integration constants as is clear. We may still decide to simplify matters choosing $b\!=\!0$ so to get (\ref{dB1}-\ref{dB4}) together with (\ref{r1}-\ref{r2}) and
\begin{eqnarray}
&R_{rtt}\!=\!ke^{\alpha}\sin{\gamma}\\
&R_{\theta tt}\!=\!-ke^{\alpha}r\cos{\gamma}\\
&R_{r\varphi t}\!=\!ke^{\alpha}\sin{\gamma}r\sin{\theta}\\
&R_{\theta\varphi t}\!=\!-ke^{\alpha}\cos{\gamma}r^{2}\sin{\theta}
\end{eqnarray}
with $k$ being the single generic integration constant.

In this second example, we have that
\begin{eqnarray}
&R_{r}\!=\!ke^{\alpha}\sin{\gamma}\!+\!(2\!+\!\partial_{\theta}\gamma)/r\\
&R_{\theta}\!=\!-ke^{\alpha}r\cos{\gamma}\!-\!r\partial_{r}\gamma\!+\!\cot{\theta}
\end{eqnarray}
\begin{eqnarray}
&B_{r}\!=\!ke^{\alpha}\cos{\gamma}\!+\!\frac{1}{r}\partial_{\theta}\alpha\\
&B_{\theta}\!=\!ke^{\alpha}r\sin{\gamma}\!-\!r\partial_{r}\alpha
\end{eqnarray}
as the components of the tensorial connection that enter into the field equations. Field equations (\ref{m1}-\ref{m2}) are thus
\begin{eqnarray}
\nonumber
&\cos{\gamma}(r\partial_{r}\gamma\!-\!\cot{\theta}\!-\!\partial_{\theta}\alpha)+\\
&+\sin{\gamma}(2\!+\!\partial_{\theta}\gamma\!+\!r\partial_{r}\alpha)\!=\!2mr
\end{eqnarray}
\begin{eqnarray}
\nonumber
&\cos{\gamma}(r\partial_{r}\alpha\!+\!2\!+\!\partial_{\theta}\gamma)-\\
&-\sin{\gamma}(r\partial_{r}\gamma\!-\!\cot{\theta}\!-\!\partial_{\theta}\alpha)\!=\!0
\end{eqnarray}
and after diagonalization
\begin{eqnarray}
&r\partial_{r}\gamma\!-\!\cot{\theta}\!-\!\partial_{\theta}\alpha\!=\!2mr\cos{\gamma}
\end{eqnarray}
\begin{eqnarray}
&2\!+\!\partial_{\theta}\gamma\!+\!r\partial_{r}\alpha\!=\!2mr\sin{\gamma}
\end{eqnarray}
which are easier to treat.

Indeed, the case $\sin{\gamma}\!=\!-1$ gives
\begin{eqnarray}
&-\cot{\theta}\!-\!\partial_{\theta}\alpha\!=\!0\\
&2\!+\!r\partial_{r}\alpha\!=\!-2mr
\end{eqnarray}
so that 
\begin{eqnarray}
e^{\alpha}r^{2}\sin{\theta}\!=\!e^{-2mr}\label{5}
\end{eqnarray}
as a \emph{fifth solution}. It is not square-integrable, but this is expected since no spinor of the form (\ref{singular}) can ever be.

Another interesting case is for $\gamma\!=\!-\theta$ giving
\begin{eqnarray}
&\cot{\theta}\!+\!\partial_{\theta}\alpha\!=\!-2mr\cos{\theta}\\
&\frac{1}{r}\!+\!\partial_{r}\alpha\!=\!-2m\sin{\theta}
\end{eqnarray}
so that 
\begin{eqnarray}
e^{\alpha}r^{2}\sin{\theta}\!=\!re^{-2mr\sin{\theta}}\label{6}
\end{eqnarray}
as a \emph{sixth solution}. Again, it is not square-integrable, but once more, this is not a problem because Majorana types of spinors, like any singular spinor, can not possibly be.

Again, notice that for the $\exp{\alpha}$ functions, compared to the $\sin{\gamma}\!=\!-1$ case (\ref{5}), the $\gamma\!=\!-\theta$ case (\ref{6}) acquires an extra $r$ dependence in the overall function and a $\sin{\theta}$ dependence in the exponential, with similar consequences for their convergence near the origin and at infinity.

We also remark the similarities between the behaviour of (\ref{1}, \ref{2}) for the modules and (\ref{5}, \ref{6}) for $\alpha$ which, in our understanding, is rather surprising, given the two different natures of $\phi$ (which describes the matter distribution) and $\alpha$ (a parameter of the tensorial connection).

The first $4$ solutions show that it is very easy to obtain solutions for the matter distribution having a non-trivial tensorial connection, but most importantly that there exist non-trivial tensorial connections. It would however be possible to have also trivial tensorial connections, which are obtained when $\gamma\!=\!-\theta$ and $\alpha\!=\!0$ identically. For this case the field equations are given according to
\begin{eqnarray}
&2Er\sin{\theta}\!+\!\partial_{\theta}\beta\!-\!2mr\sin{\theta}\cos{\beta}\!=\!0\\
&-2E\cos{\theta}\!+\!\partial_{r}\beta\!+\!2m\cos{\beta}\cos{\theta}\!=\!0\\
&-2mr\sin{\theta}\sin{\beta}\!+\!\partial_{\theta}\ln{\phi^{2}}\!=\!0\\
&2m\sin{\beta}\cos{\theta}\!+\!\partial_{r}\ln{\phi^{2}}\!=\!0
\end{eqnarray}
in general. Then it is easy to see that
\begin{eqnarray}
&\!\!\tan{\frac{\beta}{2}}\!=\!\sqrt{\frac{E-m}{E+m}}
\tan{\left(r\cos{\theta}\sqrt{E^{2}\!-\!m^{2}}\right)}\\
&\!\!\!\!\!\!\!\!\phi^{2}r^{2}\sin{\theta}\!=\!r^{2}\sin{\theta}\left[E\!+\!m\cos{\left(2r\cos{\theta}\sqrt{E^{2}\!-\!m^{2}}\right)}\right]\label{7}
\end{eqnarray}
is a \emph{seventh solution}. This solution is regular in the origin since its volume integral is finite (but not so at infinity).

We finally remark that it is not possible to obtain solutions for the matter distribution having a trivial tensorial connection for the second example. The vanishing of the tensorial connection in the first example comes from the fact that the group of relationships (\ref{aux1}-\ref{aux4}) is compatible with the vanishing of all components provided that $\alpha\!=\!0$ and $\gamma\!=\!-\theta$ but the vanishing of the tensorial connection in the second example requires (\ref{bux1}-\ref{bux2}) to be compatible with the vanishing of all components, meaning
\begin{eqnarray}
&-r|\!\sin{\theta}|^{2}\!=\!0\\
&-r^{2}\sin{\theta}\cos{\theta}\!=\!0
\end{eqnarray}
and this cannot be done. Therefore, in the second example the tensorial connection can not be vanished.
\section{Considerations}
In the previous section \ref{C} we have provided seven exact solutions of the Dirac equations, four for regular spinors with non-trivial tensorial connection, two for the singular spinor of Majorana type still with a non-trivial tensorial connection. A final one for regular spinors with the trivial tensorial connection, commenting that a trivial tensorial connection is not possible for the singular spinors.

Let us now compare the first six solutions, so to extract some general information about the tensorial connection.

In the first example, we had two pairs of sub-cases, one given when $\sin{\gamma}\!=\!-1$ so that in particular
\begin{eqnarray}
&R_{rtt}\!=\!-2\varepsilon\sinh{\alpha}\\
&R_{r\varphi t}\!=\!-2\varepsilon r\sin{\theta}\cosh{\alpha}
\end{eqnarray}
and the other given by $\gamma\!=\!-\theta$ yielding
\begin{eqnarray}
&R_{rtt}\!=\!-2\varepsilon\sin{\theta}\sinh{\alpha}\\
&R_{\theta tt}\!=\!-2\varepsilon r\cos{\theta}\sinh{\alpha}\\
&R_{r\varphi t}\!=\!-2\varepsilon r|\!\sin{\theta}|^{2}\cosh{\alpha}\\
&R_{\theta\varphi t}\!=\!-2\varepsilon r^{2}\sin{\theta}\cos{\theta}\cosh{\alpha}
\end{eqnarray}
while in the second example, we had two sub-cases, given when $\sin{\gamma}\!=\!-1$ so that
\begin{eqnarray}
&R_{rtt}\!=\!-ke^{\alpha}\\
&R_{r\varphi t}\!=\!-ke^{\alpha}r\sin{\theta}
\end{eqnarray}
and for $\gamma\!=\!-\theta$ furnishing
\begin{eqnarray}
&R_{rtt}\!=\!-ke^{\alpha}\sin{\theta}\\
&R_{\theta tt}\!=\!-ke^{\alpha}r\cos{\theta}\\
&R_{r\varphi t}\!=\!-ke^{\alpha}r|\!\sin{\theta}|^{2}\\
&R_{\theta\varphi t}\!=\!-ke^{\alpha}r^{2}\sin{\theta}\cos{\theta}
\end{eqnarray}
where obvious similarities emerge. Most notably, the first example in the limit of large $\alpha$ is just the second example, provided that constants $\varepsilon$ and $k$ are renamed the same.

We notice that the component $R_{r\theta\theta}$ disappears in the $\gamma\!=\!-\theta$ cases compared to the $\sin{\gamma}\!=\!-1$ cases, and that this is the only difference between these two sets of cases.

It is worth remarking that components $R_{r\varphi\varphi}$ and $R_{\theta\varphi\varphi}$ are common to all cases as they are purely geometrical.

An important point is that not all components of the tensorial connection vanish at infinity (even in the cases in which the spinor field is square-integrable). However, this fact is of little concern, since a tensorial connection, albeit real, is not related to any source, it does not have a propagation, and its non-locality poses no threat.
\section{Extended QFT}
Thus far, we have established the general mathematical theory of spinor fields in their polar form demonstrating how tensorial connections naturally emerge. Moreover it is also possible to see that specific tensorial connections are present for Hydrogen Atoms \cite{Fabbri:2018crr}. Mathematically far from being ill-defined, and physically far from being not applicable in important systems, it is therefore surprising that the $R_{ij\mu}$ tensors occupy a marginal place in QFT.

The reason lies in the fact that for plane waves, $R_{ij\mu}$ is proven to vanish identically. The way to see that is by considering plane waves, which are defined as the specific type of spinors verifying the condition
\begin{eqnarray}
&i\boldsymbol{\nabla}_{\mu}\psi\!=\!P_{\mu}\psi
\end{eqnarray}
which shows, upon comparison against (\ref{decspinder}), that
\begin{eqnarray}
&(-\frac{i}{2}\nabla_{\mu}\beta\boldsymbol{\pi}
\!+\!\nabla_{\mu}\ln{\phi}\mathbb{I}\!-\!\frac{1}{2}R_{ij\mu}\boldsymbol{\sigma}^{ij})\psi\!=\!0
\end{eqnarray}
must hold. Thus the linear independence of the Clifford matrices implies that $R_{ij\mu}\!\equiv\!0$ beside the restrictions for the degrees of freedom $\nabla_{\nu}\beta\!\equiv\!0$ and $\nabla_{\mu}\phi\!\equiv\!0$ in general.

Hence, $R_{ij\mu}\!\equiv\!0$ is the consequence of the contrived nature of QFT, specifically in the use of the plane waves to describe the propagation of point-like particles.

It is also possible to prove that the tensorial connection can encode the information encoded by the momentum.

To see that, consider the polar form (\ref{regular}). In it, the $\boldsymbol{L}$ term contains a global phase that is linked, through (\ref{P}), to the momentum. The global phase, however, acting on the constant spinor, which is an eigen-state for rotations around the third axis,\! comes to be equivalent to a rotation around the third axis itself. Such rotation is encoded in $\boldsymbol{L}$ and so, through (\ref{R}), to the third-axis component of the tensorial connection. Then the structure of (\ref{LdL}) shows a connection between global phase and third-axis rotation in the sense that the two are formally interchangeable.

This can of course be translated into the structures of momentum and tensorial connection. And specifically, a given momentum $P_{\mu}$ can always be written as the corresponding tensorial connection $-2P_{\mu}\!=\!R_{12\mu}$ or
\begin{eqnarray}
&\!-\!2P_{\mu}u^{a}s^{b}\varepsilon_{abij}\!=\!\Delta R_{ij\mu}
\end{eqnarray}
in manifestly covariant form. This means that re-naming
\begin{eqnarray}
&R_{ij\mu}\!-\!2P_{\mu}u^{a}s^{b}\varepsilon_{abij}\!=\!R'_{ij\mu}\\
&P'_{\mu}\!=\!0
\end{eqnarray}
so that also
\begin{eqnarray}
&\nabla_{\mu}u_{i}\!=\!R'_{ji\mu}u^{j}\\
&\nabla_{\mu}s_{i}\!=\!R'_{ji\mu}s^{j}
\end{eqnarray}
entails no loss of generality. Thus momenta can always be re-absorbed within tensorial connections in general cases.

We shall see later what happens in the complementary case in which no momentum can be defined at all.
\section{Neutrality Condition}
To a first sight, discussing about the possibility to have no momentum in QFT looks harsh, since the full setting of QFT lies on the hypothesis of plane waves, or more precisely on the fact that it is possible to perform a Fourier transformation to map fields in momentum space. How then can be possible to have no momentum? The answer, though very unexpected, is however rather obvious, and that is, in performing a Fourier transformation, we treat particles as if they had a single momentum. This may be true for some, but for particles represented by the spinor fields, which have two chiral parts, this assumption might be too restrictive. There is, in principle, still no problem for Dirac particles, since the two irreducible components have the same helicity. But Majorana particles, with opposite helicity states, do suffer. Indeed, Majorana fields need more than just the momentum, to have a consistent definition of covariant derivative. Let us see it in detail.

As is clear from the covariant derivative, the Majorana spinors have no contribution coming from $\alpha$ and so
\begin{eqnarray}
&\boldsymbol{\nabla}_{\mu}\psi
\!=\!(-\frac{1}{2}R^{ij}_{\phantom{ij}\mu}\boldsymbol{\sigma}_{ij}\!-\!iP_{\mu}\mathbb{I})\psi
\label{Majorana}
\end{eqnarray}
with only tensorial connection and momentum. Because $\boldsymbol{\gamma}^{2}\boldsymbol{\gamma}_{i}^{*}\boldsymbol{\gamma}^{2}\!=\!\boldsymbol{\gamma}_{i}$ we would then have that
\begin{eqnarray}
&\boldsymbol{\nabla}_{\mu}(i\boldsymbol{\gamma}^{2}\psi^{*})
\!=\!(-\frac{1}{2}R^{ij}_{\phantom{ij}\mu}\boldsymbol{\sigma}_{ij}
\!+\!iP_{\mu}\mathbb{I})(i\boldsymbol{\gamma}^{2}\psi^{*})
\end{eqnarray}
but from the Majorana condition $i\boldsymbol{\gamma}^{2}\psi^{*}\!=\!\psi$ we get
\begin{eqnarray}
&\boldsymbol{\nabla}_{\mu}\psi
\!=\!(-\frac{1}{2}R^{ij}_{\phantom{ij}\mu}\boldsymbol{\sigma}_{ij}
\!+\!iP_{\mu}\mathbb{I})\psi
\end{eqnarray}
to be compared against (\ref{Majorana}). Therefore $P_{\mu}\!=\!0$ and thus
\begin{eqnarray}
&\boldsymbol{\nabla}_{\mu}\psi\!=\!-\frac{1}{2}R^{ij}_{\phantom{ij}\mu}\boldsymbol{\sigma}_{ij}\psi
\end{eqnarray}
which is just the expression (\ref{m}). Therefore no Majorana spinor can ever be characterized by anything more than the tensorial connection in its covariant derivative.

At a dynamical level, this can be assessed by having a look at what happens when field equations (\ref{m1}-\ref{m2}) are taken with no tensorial connection. One of them, that is (\ref{m2}), reduces to $mU^{\sigma}\!=\!0$ which is solved by $m\!=\!0$ solely.

Consequently, the Majorana spinor is a case for which the tensorial connection \emph{can not} ever be trivial.
\section{ELKO}
This situation calls our attention on a discussion on the role of Majorana-like spinors in physics, which has been carried on in recent times by Ahluwalia and co-workers in \cite{Ahluwalia:2004ab, Ahluwalia:2004sz, Ahluwalia:2016jwz, Ahluwalia:2016rwl}. The pivotal point of such a discussion is on the fact that Majorana-like spinors, that is spinors which are eigen-states of charge-conjugation, do not match all features spinors should have, and in particular these spinors cannot solve the massive Dirac equation. Ahluwalia's solution for this situation is to acknowledge the reluctance of such spinors in being solutions of the first-order Dirac equation requiring them to be solutions only of second-order Klein-Gordon field equations. This property would provide these spinors, known as ELKO, with the unitary mass-dimension, making them dark matter candidates.

The general reasoning is that the conditions of charge conjugation (being self- or antiself-conjugated) can (each) be realized by two opposite helicity states $\lambda_{+}$ and $\lambda_{-}$ as they are normally called. Hence the Dirac operator acts as $P\!\!\!\!/\lambda_{+}\!=\!im\lambda_{-}$ and $P\!\!\!\!/\lambda_{-}\!=\!-im\lambda_{+}$ and therefore no Dirac equation in momentum space is valid. Nevertheless, this argument stems from the fact that we generally have
\begin{eqnarray}
&\boldsymbol{\gamma}^{0}\lambda_{+}\!=\!i\lambda_{-}
\end{eqnarray}
which cannot be written in the form of eigen-state of $\boldsymbol{\gamma}^{0}$ as clear. However, the two helicities are also related by 
\begin{eqnarray}
&\boldsymbol{\gamma}^{1}\boldsymbol{\gamma}^{3}\lambda_{+}\!=\!-\lambda_{-}
\end{eqnarray}
and therefore we have that
\begin{eqnarray}
&\boldsymbol{\gamma}^{0}\boldsymbol{\gamma}^{1}\boldsymbol{\gamma}^{3}\lambda_{+}\!=\!i\lambda_{+}
\end{eqnarray}
which can indeed be written as eigen-state. In fact
\begin{eqnarray}
&i\boldsymbol{\gamma}^{0}(-\frac{1}{2}4m\boldsymbol{\sigma}^{13}\lambda_{+})\!=\!m\lambda_{+}
\end{eqnarray}
as easy to see. Then, provided that we have
\begin{eqnarray}
&2m\!=\!R_{130}
\end{eqnarray}
we can finally write
\begin{eqnarray}
&i\boldsymbol{\gamma}^{\nu}(-\frac{1}{2}R_{ab\nu}\boldsymbol{\sigma}^{ab}\lambda_{+})\!=\!m\lambda_{+}
\end{eqnarray}
identically. Because of the definition of derivative of Majorana spinors (\ref{m}), the above is therefore the usual Dirac equation for the $\lambda_{+}$ spinor. The same would be obtained for the  $\lambda_{-}$ spinor, so that we can write
\begin{eqnarray}
&i\boldsymbol{\gamma}^{\nu}\boldsymbol{\nabla}_{\nu}\lambda\!=\!m\lambda
\end{eqnarray}
in the most general circumstance whatsoever.

So as stated in \cite{Ahluwalia:2004ab, Ahluwalia:2004sz, Ahluwalia:2016jwz, Ahluwalia:2016rwl}, ELKO in their pure momentum form cannot be eigen-states of the Dirac equation. Still, the problem is not the Dirac equation, but the fact that the Majorana spinors are written in the pure momentum form. In fact the pure momentum form of the Majorana spinor is too restrictive to convey the double helicity that pertains to these fields. Or to put it differently, one must have the possibility to write one Fourier transformation for each of the two independent helicities. This enriched structure can be given by the tensorial connection, as we have seen above. And for enlarged underlying space-time structures the Majorana spinor \emph{are} solution of the Dirac equations, as the second example of section \ref{C} showed.
\section{Conclusion}
In this paper, we have presented seven exact solutions of the Dirac equations: for regular spinors, a specific tensorial connection gave us a first solution (\ref{1}) which was already known \cite{Fabbri:2019kfr}, and which we have now accompanied by an alternative form in (\ref{3}), another alternative form in (\ref{2}), and another in (\ref{4}).\! We have repeated such an analysis for singular spinors of Majorana type, where the second example of tensorial connection has allowed us to obtain solution (\ref{5}), alongside to (\ref{6}). For the regular spinors, we saw that the zero tensorial connection can be obtained and it gave a seventh solution (\ref{7}). A singular spinor of Majorana type however can not have any trivial tensorial connection for which a solution can be found.

Apart from the solution that was already known, none of the other solutions converges at infinity, although they are all regular at the origin. That they cannot be physical in an open domain does not decrease their interest for an application to closed and limited domains, or for a purely topological perspective. And in any case, much insight is likely to be gained from their comparative analysis.

Solutions (\ref{5}-\ref{6}) are one example of tensorial connection ensuring an exact solution for Majorana spinors.

And even in the case of regular spinors those pertaining to the non-trivial tensorial connections, given by (\ref{1}), (\ref{2}), (\ref{3}), (\ref{4}), behave better than the one arising from a trivially null tensorial connection, given by (\ref{7}).

This comparative analysis seems to suggest that there indeed are cases of spinors that necessitate the tensorial connection in order to be well defined, and cases that do not really need it but which would still be better behaved whenever some specific tensorial connection is present.

More insight in the role of the tensorial connection can only be obtained by finding more special solutions.

A further work is already being undertaken.

\

Manuscript has no associated data in repositories.

\

The author declares no conflict of interest.

\

This work received no funding.

\end{document}